# Novel rhenium carbides at 200 GPa

Saiana Khandarkhaeva[1*], Timofey Fedotenko[2], Maxim Bykov[3], Elena Bykova[3], Stella Chariton[4], Pavel Sedmak[5], Konstantin Glazyrin[6], Vitali Prakapenka[4], Natalia Dubrovinskaia[2,7] and Leonid Dubrovinsky[1]

[1] *Bayerisches Geoinstitut, University of Bayreuth, Universitätstraße 30, 95440 Bayreuth, Germany*

[2] *Material Physics and Technology at Extreme Conditions, Laboratory of Crystallography University of Bayreuth, Universitätstraße 30, 95440 Bayreuth, Germany*

[3] *Geophysical Laboratory, Carnegie Institution of Washington, 5251 Broad Branch Road NW, 20015 Washington, District of Columbia, USA*

[4] *Center for Advanced Radiation Sources, University of Chicago, 5640 S. Ellis, 60637 Chicago, Illinois, USA*

[5] *European Synchrotron Radiation Facility, BP 220, 38043 Grenoble Cedex, France*

[6] *Photon Science, Deutsches Elektronen-Synchrotron, Notkestraße 85, 22607 Hamburg, Germany*

[7] *Theoretical Physics Division, Department of Physics, Chemistry and Biology (IFM) Linköping University, SE-581 83, Linköping, Sweden*

*Corresponding author: Saiana.Khandarkhaeva@uni-bayreuth.de

**Abstract:** Laser heating of rhenium in a diamond anvil cell to 3000 K at about 200 GPa results in formation of two previously unknown rhenium carbides, hexagonal WC-type structured ReC and orthorhombic $TiSi_2$-type structured $ReC_2$. The Re-C solid solution formed at multimegabar pressure has the carbon content of ~20 at%. Unexpectedly long C-C distances (~1.76-1.85 Å) in "graphene-like" carbon nets in the structure of $ReC_2$ cannot be explained by a simple covalent bonding between carbon atoms and suggest that at very high pressures the mechanism of interaction between carbon atoms in inorganic compounds may be different from that considered so far.

Chemical compounds of 5d transition metals and carbon or other first-row elements, as B and N, often possess interesting properties attributed to strong covalent bonding.[1] Many of carbides, borides, and nitrides reveal very high melting points (for example, over 3500 K for ZrC, NbC, HfC, TaC),[2] large bulk moduli ($K_0$>390 GPa for of $Re_2C$,[3] $Re_2N$,[4] $ReN_2$,[5]

IrN$_2$,[6] OsB,[7] Os$_2$B$_3$,[7] OsB$_2$[7]), and very high hardness (H$_v$>35 GPa for ReN$_2$,[5] WB$_4$[7]). Transition-metal carbides belong to a large group of industrially important materials.

The rhenium-carbon system provides a striking example of the pressure effect on elements reactivity and the binary phase diagram. At ambient pressure, rhenium does not form stoichiometric carbides; carbon dissolves into rhenium up to 28.45 at% at the eutectic temperature (2778 K).[8] However, even very moderate pressure, just above 6 GPa (and high temperatures) was reported to promote formation of a Re-C compound.[9,10] Its correct chemical composition (Re$_2$C) and crystal structure of anti-MoS$_2$ type (hexagonal primitive, *hP*, space group *P*6$_3$/mmc) were established relatively recently on the basis of X-ray powder diffraction, Raman spectroscopy data and DFT calculations.[11,12] No other stoichiometric carbides apart of *hP*-Re$_2$C have been observed at pressures up to ~70 GPa and temperatures ~4000 K.[3]

The *hP*-Re$_2$C was found to be isostructural with Re$_2$N.[11] This analogy and the recently observed very complex and unexpected behavior of the Re-N system, featuring numerous nitrogen-rich compounds,[5,13] stimulated the study of potential reactions between Re and C at multimegabar pressures. Awareness of these reactions is also of a primary interest for the development of the methodology of ultra-high pressure high temperature experiments, in which Re gaskets are commonly used. The range of currently achievable static pressures has been extended to ~1000 GPa due to implementation of double-stage diamond anvil cells (dsDAC) and to ~600 GPa with toroidal type anvils (tDAC).[14–17] In order to achieve such extreme pressures, the linear size of samples and sample chambers should be drastically decreased. At pressures above ~150 GPa, a pressure chamber's diameter (made, as a rule, of Re) is usually smaller than 50 μm, and in dsDACs above 300 GPa it is less than 10 μm. Meanwhile, the size of a laser beam in typical laser heating (LH) setups used in DAC experiments varies from 15 to 50 μm at FWHM.[18,19] As a result, irradiation of, at least the edge of a Re-gasket, by the laser beam during laser heating becomes unavoidable and may lead to a chemical reaction between Re and carbon of the diamond anvils. Therefore, correct interpretation of the results of experiments with laser-heating at ultra-high static pressures requires knowledge about possible products of rhenium and carbon interaction.

Here, we report on the *in situ* study of Re-C compounds formed due to chemical interactions between diamond anvils and the rhenium gasket after pulsed laser heating in DACs at about 200 GPa. The structures of all of the synthesized rhenium carbides, Re$_2$C, ReC$_2$, ReC, and ReC$_{0.2}$ were solved and refined using single-crystal X-ray diffraction (SCXRD) providing direct and unequivocal data to judge on both the atomic arrangement and chemical composition of the crystalline matter.

Two DACs have been prepared, dsDAC for Experiment №1 and conventional assembly for Experiment №2. Secondary anvils for Experiment №1 were made from nano-crystalline diamond (NCD) spheres of about 15-20 μm in diameter. The dsDAC in Experiment №1 was compressed up to ~415 GPa (according to the diamond Raman shift on the secondary anvil),[20] then the sample (Re flake) was laser heated until the first bright flash of light that held for less than a second (temperature was not measured). After the first heating attempt pressure dropped down to ~200 GPa probably due to failure of the secondary anvils. However, the DAC remained intact and rhenium in the central area was again carefully laser heated in a pulsed mode (1 μs pulses, 25 kHz repetition rate, and maximum temperature of about 3000 K).[21] Pressure did not change upon the repeated laser heating (Supporting Information, Figure S1) and the DAC with the temperature-quenched material was investigated using powder and single crystal X-ray diffraction (for details see the Supporting Information). A 2D diffraction map collected across the sample chamber revealed the presence of not only Re, but four additional phases (Figure 1). All of them have been identified and are described in detail below.

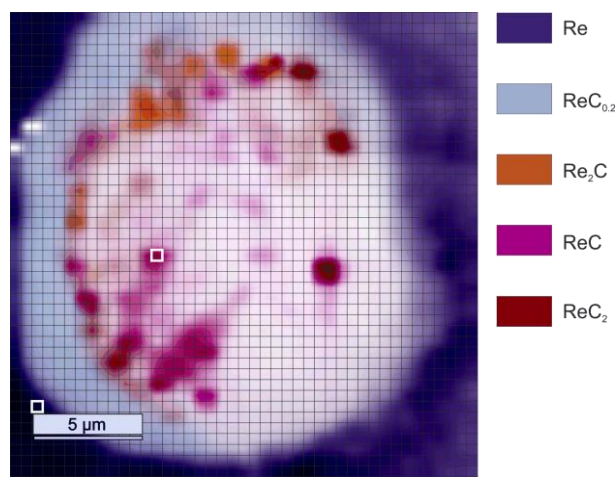

*Figure 1.* 2D map of X-ray diffraction patterns. The color intensity is proportional to the intensity of the following reflections: the (100) and (101) reflections of the Re gasket for the violet region; the (100) and (101) reflections of $ReC_{0.2}$ for the light blueish-gray region; the (105) reflection of $Re_2C$ for the orange region; the (101) reflection of ReC with the WC-type structure for the purple region; the (002), (111), (020) reflections of $ReC_2$ for the dark red region. Characteristic diffraction images highlighted with white rectangles are shown in Figure S2.

The XRD patterns collected from the edge of the Re-gasket show typical continuous powder diffraction rings of rhenium (Supporting Information, Figure S2a) not exposed to the laser beam. The unit cell volume of Re determined from the Le Bail fit varies from 22.000(5) Å$^3$ to 21.978(7) Å$^3$ for different patterns (Supporting Information, Figure S3). This corresponds to pressures of ~230-240 GPa according to equation of state (EOS) from Dubrovinsky et al.,[14] or ~190-195 GPa, according to Anzellini et al.[22] (Supporting Information, Table S1).

One of the crystalline phases we found within the laser-heated area is the hexagonal $hP$-$Re_2C$ ($P6_3/mmc$, № 194) (Figure 2a).[11,12,23] The parameters of the unit cell were found to be $a$=2.5860(9) Å, $c$=9.272(3) Å, V=53.70(3) Å$^3$ (Supporting Information, Table S2, Figure S4a). As written above, in this experiment the pressure, as determined from the Raman shift of the diamond anvil (Supporting Information, Figure S1), was of about 200 GPa, whereas the Re EOS gave up to 240 GPa. These values do not match the pressure of 172(13) GPa, determined for the given unit cell volume of $hP$-$Re_2C$ according to the EOS reported by Juarez-Arellano et al.[3] The EOS of $Re_2C$ in ref.[3] was determined on the basis of powder XRD from a sample in hard pressure transmitting medium that led to a significant uncertainty in the bulk modulus: K=405(30) GPa (K´=4.6).[3] Comparison of our data with literature motivated us to make an independent measurement to establish the EOS of $hP$-$Re_2C$ on the basis of SCXRD. In the experiment described in Supporting Information, Experimental Procedures, the P-V data were obtained up to 50 GPa from a single crystal of $hP$-$Re_2C$ pressurized in a soft (Ne) pressure transmitting medium (Supporting Information, Table S3, Figure S5). The parameters of the 3$^{rd}$ order Birch-Murnaghan equation of state of $hP$-$Re_2C$ ($V_0$=69.18(4) Å$^3$/unit cell, K=375(15) GPa, K´=5.0(1)) we obtained only slightly (within uncertainties) differ from the values reported by Juarez-Arellano et al.[3] According to our EOS $hP$-$Re_2C$ synthesized in Experiment №1 was under the pressure of 180(7) GPa (Figure 3a). The structure of $hP$-$Re_2C$ (Figure 2a, Supporting Information, Table S2) determined from SCXRD is in a good agreement with model described by Friedrich et al.[11] It is characterized by the stacking sequence AABB of layers of Re atoms (with four Re atoms per a unit cell). Carbon atoms occupy the *2d* Wyckoff position in trigonal prisms formed by Re atoms with the Re-C distances of 1.997(1) Å (Figure 2e).

The structure of another phase, identified as ReC, has a hexagonal unit cell with the lattice parameters $a$=2.5510(9) Å, $c$=2.7048(11) Å, and $V$=15.24(1) Å$^3$ (Figure 1b) (Supporting Information, Table S2, Figure S4b). The structure solution and refinement revealed that $hP$-ReC belongs to the WC structure type (*P-6m2*, №187) with characteristic *c/a* ratio (~0.94). Like in $Re_2C$, carbon atoms are located in trigonal prisms formed by Re atoms with the Re-C distances equal to 2.000(2) Å (Figure 2f).

One more carbide, $ReC_2$, with the orthorhombic structure found in Experiment №1 (Figure 2c, Supporting Information, Table S2, Figure S4c) has the lattice parameters $a$=3.3367(10) Å, $b$=4.3155(16) Å, $c$=5.6220(13) Å, and V=80.95(4) Å$^3$, and a space group *Fmmm* (№69). Remarkably, the same orthorhombic phase ($a$=3.2880(9) Å, $b$=4.2088(9) Å, $c$=5.5645(8) Å, and V= 77.00(3) Å$^3$) was found in Experiment №2, in which iron oxide (FeO) in a Ne pressure medium was compressed to 219(5) GPa and pulsed-laser heated up to ~2500 K.[19] The relatively large beam (of about 25 μm at FWHM) irradiated the Re gasket used in Experiment №2 and ReC$_2$ was found at the border of the pressure chamber.

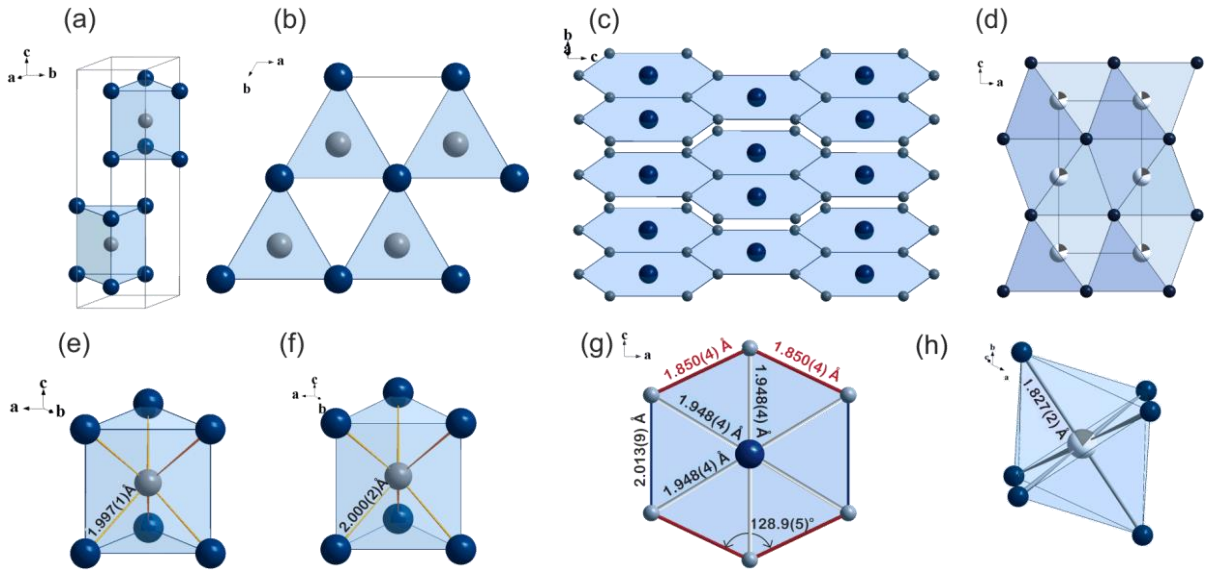

*Figure 2*. Crystal structures of Re-C compounds observed in Experiment №1. (a) hP-Re$_2$C, (b) WC-type structured ReC, (c) TiSi$_2$-type structured oF-ReC$_2$; and (d) B8-type structured interstitial solid solution ReC$_{0.2}$. Dark blue balls designate Re atoms; dark gray balls – carbon atoms; white balls with a dark gray sector symbolize a partial occupancy of the position by carbon atoms. Carbon coordination polyhedra in the crystal structures of Re$_2$C (e) and ReC (f) are strikingly similar with the Re-C bond length ~2 Å (shown by yellow color). (g) A single hexagon in the graphene-like web of carbon atoms, which coordinate each Re atom in the structure of ReC$_2$. The shorter C-C distances of 1.850(4) Å are highlighted by red color. (h) A single octahedron formed by Re atoms in the structure of ReC$_{0.2}$, which incorporates carbon atoms with the occupancy of 20 %. The Re-C distances are relatively short (1.827(2) Å).

The structure solution and refinement of the orthorhombic phase, in both Experiment №1 and №2, resulted in the chemical composition ReC$_2$ and the structure as shown in Figure 2c. Carbon atoms form graphene-like nets in the *ac*-plane stacking along the *b* direction. The hexagons are not ideal (Figure 2g), they have two longer and four shorter sides. The shorter ones, found in Experiment №1 and №2, have the lengths of 1.850(4) Å and 1.758(3) Å, respectively. These C-C distances are significantly larger than those in graphene (or graphite) (about 1.42 Å at ambient conditions).[24] Rhenium atoms are located in the center of each of carbon hexagons (Figure 2g) and thus each Re has 6 closest carbon neighbors at a distances of ~1.75-1.95 Å.

Carbon-rhenium layers stack along *b*-axis with a translation (½, 0, ½) (Figure 2c). Whereas the Re-C layers are obviously helpful to give a clear geometrical presentation of the *oF*-ReC$_2$ structure, the structure is not "layered", as the shortest Re-Re distances between the "layers" (~2.72 Å) are similar to those in *hP*-Re$_2$C (~2.5-2.6 Å) and *hP*-ReC (~2.55-2.70 Å) at the same pressure. However, a strong diffuse scattering, especially evident in the reciprocal space, suggests a disorder in the stacking of Re-C "layers" along the *b*-axis – (Supporting Information, Figure S6a).

Relatively long C-C contacts in *oF*-ReC$_2$ (~1.76-1.85 Å that is about 30% longer than in sp$^2$-, or 20% longer than in sp$^3$-bonded carbon) indicate that usual carbon-carbon covalent interaction in this compound unlikely exists. Indeed, *oF*-ReC$_2$ is isostructural to one of the polymorphs of titanium disilicide (TiSi$_2$) with a pseudo-C11$_b$-type of structure (space group *Fmmm*, №. 69), a=4.428 Å, b=4.779 Å, c=9.078 Å, V=192.1 Å$^3$).[25] The shortest Si-Si distance in *oF*-TiSi$_2$ at ambient conditions is ~2.57 Å (compare to ~2.35 Å in diamond-structured Si),[26] which is close to ~2.54 Å in C54-type structured *oF*-TiSi$_2$ and to ~2.53 Å in a Zintl phase CaSi. For the last two cases, *ab initio* simulations suggest a significant covalent interaction between Si atoms in zig-zag chains geometrically similar to the C-C chains in our *oF*-ReC$_2$ carbide.[27,28] At the same time, the shortest C-C distances in *hP*-ReC and *hP*-Re$_2$C at 180(7) GPa (in this study, Experiment №1) are 2.5510(9) Å and 2.5860(9) Å, correspondingly, and these values agree with those known for transition metal carbides containing isolated carbon atoms (~2.8 Å and above at ambient pressure). Thus, the nature of chemical bonding in *oF*-ReC$_2$ is probably different from that in other rhenium carbides - *hP*-ReC and *hP*-Re$_2$C.

The phase map built on the basis of powder XRD data (Figure 1) shows that several carbides (Re$_2$C, ReC, and ReC$_2$) formed within the pressure chamber, and one more phase has been synthesized at its periphery, at the border with the Re gasket. The diffraction patterns are mostly characterized by powder rings, but in a few points in the map it was possible to collect single-crystal data sets (Supporting Information, Figure S4d). Indexing of the powder and single-crystal data resulted in a hexagonal unit cell with the lattice parameters *a*=2.6028(12) Å, *c*=4.161(2) Å, *V*= 24.41(2) Å$^3$, and a space group *P6$_3$/mmc* (№.194) (Supporting Information, FigureS4d, Table S2). This may be interpreted as Re at about 100 GPa, according to the EOS of Anzellini et al.[22] or at about 115 GPa, according to the EOS of Dubrovinsky et al.[14] Indeed, single-crystal data suggest that rhenium atoms form hexagonal closed packing as expected for Re. However, the electron density, localized in the octahedral voids of the *hcp*-Re, is arranged like in the B8 (NiAs)-type structure (Figure 2d) and suggests this phase to be a Re-C interstitial solid solution based on the B8-type structure (Figure 2d, h). Single-crystal diffraction data at hands are not sufficient to refine the carbon atoms occupancies, especially in the case of such a huge difference in X-ray scattering factors of Re and C. Fortunately, we noticed that the unit cell volumes per atom for $Re_xC_y$ compounds (Re,[14] $hP$-$Re_2C$, $hP$-ReC, $oF$-$ReC_2$, and C (diamond)[29]) at 180(7) GPa, if plotted as a function of carbon content, all appear along the common straight line (Figure 3b). Considering the volume of the B8 rhenium-carbon solid solution, we have estimated its composition as $ReC_{0.2}$, which shows that multimegabar pressures do no increase carbon solubility in rhenium, compared to previously observed compositions.[8]

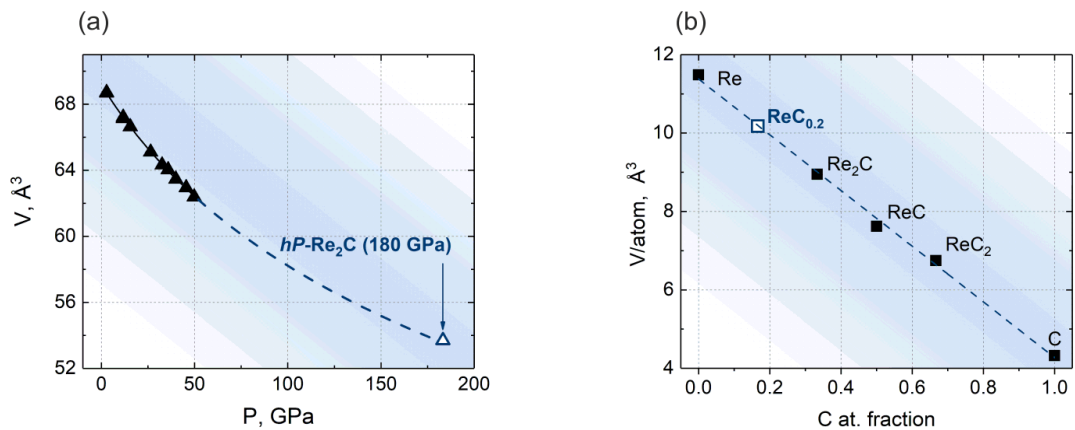

*Figure 3.* The unit cell volume of $hP$-$Re_2C$ as a function of pressure (a) and the dependence of the unit cell volumes per atom on the atomic fraction of carbon for various $Re_xC_y$ compounds at 180(7) GPa (b). Black triangles are experimental data points obtained using SCXRD on compression of $hP$-$Re_2C$ up to ~50 GPa in a soft (Ne) pressure transmitting medium. Solid black line is a fit of the P-V experimental data using the 3$^{rd}$ order Birch-Murnaghan EOS ($V_0$=69.18(4) Å$^3$/unit cell, K=375(15) GPa, K´=5.0(1)) and the dashed line is its extrapolation. Open triangle (corresponds to the volume obtained for $hP$-$Re_2C$ in one pressure point in Experiment №1 (see text). Black squares are experimental data points for observed Re-C compounds; open square designates $ReC_{0.2}$, in which the atomic fraction of carbon was determined in accordance with this empiric 'V/atom vs C at. Fraction' relationship.

A comparison of our data for Re-C compounds and recent reports on the Re-N system demonstrates obvious crystal-chemical similarities in the two systems for compositions Re:C≥1:1. Two carbides, $Re_2C$ and ReC, are isostructural to $Re_2N$ and ReN. They are built up of $CRe_6$ or $NRe_6$ trigonal prisms, and even Re-C and Re-N interatomic distances at the same pressures are very similar.[4,5] Contrary, the structures of $ReC_2$ and $ReN_2$[5] have nothing in common and there are no signs of formation of rhenium polycarbides with more than two carbon atoms per a formula unit, unlike to the rhenium-nitrogen system.[13] Possible explanation of such differences may be a tendency of nitrogen to form di-nitrogen and poly-nitrogen anions (based on N4 units[13,30–32]) at high-pressures and high-temperatures (HPHT), whereas rhenium carbides do not reveal formation of polycarbon anions at such conditions. There is a number of experimental works and theoretical predictions that suggest progressive polymerization of carbon atoms at high pressures.[33–38] In all cases polymerization results in formation of short (~1.6 Å or shorter) C-C bonds for single bonded carbon atoms. For hexagonal carbides $hP$-Re$_2$C or $hP$-ReC at ~200 GPa, the C-C distances are longer than 2.5 Å, which suggest the absence of chemical bonding between carbon atoms in the crystal structure. In case of rhenium dicarbide, $oF$-ReC$_2$, we observed formation of graphene-like' carbon layers, in which the C-C distances (~1.76-1.85 Å) are much longer than expected for sp$^2$- or sp$^3$-bonded carbon atoms in inorganic compounds; still, in alkanes the C-C bond length as long as 1.704 Å has been detected.[39] It suggests that at very high pressures a mechanism of interaction between carbon atoms in inorganic compounds may be different from that considered so far.

To summarize, we have extended the knowledge about the chemical interaction between rhenium and carbon to ~200 GPa. Two novel carbon-rich rhenium carbides – WC-type structured $hP$-ReC and TiSi$_2$-type structured $oF$-ReC$_2$ – were synthesized (hitherto only Re$_2$C was known in the Re-C system). The fact that rhenium and carbon can produce numerous compounds at HPHT conditions should be taken into account upon planning of experiments in LHDACs at multimegabar pressures.

**Acknowledgements**

N.D. and L.D. thank the Federal Ministry of Education and Research, Germany (BMBF, grants No. 5K16WC1 and No. 05K19WC1) and the Deutsche Forschungsgemeinschaft (DFG projects DU 954-11/1, DU 393-9/2, and DU 393-13/1) for financial support. N.D. thanks the Swedish Government Strategic Research Area in Materials Science on Functional Materials at Linköping University (Faculty Grant SFO-Mat-LiU No. 2009 00971). We acknowledge DESY (Hamburg, Germany), a member of the Helmholtz Association HGF, for the provision of experimental facilities. Parts of this research were carried out at Petra III and we would like to thank Dr. H.-P. Liermann for assistance in using photon beamline P02.2. We thank Dr. Alexander Kurnosov (Bayerisches Geoinstitut, Bayreuth) for assistance during the preparation of DACs.

Detailed information on the crystal data obtained from single-crystal XRD refinement of the observed rhenium carbides at 180(7) GPa is provided in the Supporting Information, Table S2. CSD 1976969, 1976970, 1976971, 1976973 contain the supplementary crystallographic data of $Re_2C$, $ReC$, $ReC_{0.2}$ and $ReC_2$ correspondently. These data can be obtained from The Cambridge Crystallographic Data Centre.